\documentclass[preprint,epsf]{aastex}
\usepackage{amssymb,latexsym,graphics,eufrak}

\def\a0{a_0}
\def\msun{M_{\odot}}
\def\kms{{\rm km~ s}^{-1}~}
\def\cmst{{\rm cm~ s}^{-2}~}

\def\Halpha{H$_{\alpha}~$}
\def\deg{^o}
\def\kpc{{\rm kpc~}}
\def\mpc{{\rm Mpc~}}

\begin{document}

\title{{\bf MOND rotation curves of very low mass spiral galaxies}}
\author{Mordehai Milgrom$^1$ and Robert H. Sanders$^2$}
\affil{ $^1$Center for Astrophysics, Weizmann
Institute\\
 $^2$Kapteyn Astronomical Institute, Groningen, NL}

\begin{abstract}
We present MOND analysis for several of the lowest mass disc
galaxies currently amenable to such analysis--with (baryonic) masses
below $4\cdot 10^8\msun$. The agreement is good, extending the
validity of MOND and its predicted mass velocity relation, to such
low masses.
\end{abstract}


\section{Introduction}
The MOND paradigm \cite{mil83} has received a substantial boost
with the recent advent of relativistic formulations of the theory
[e.g., bekenstein(2004)].  MOND's {\it raison d'etre} remains,
however, largely phenomenological. It is crucial then to continue
to test the predictions of MOND in different contexts. The
flagship of phenomenological testing of MOND is the analysis of
rotation curves of disc galaxies because such data provides the
most precise description of the radial dependence of force in an
extragalactic context. Indeed there is a large body of work
applying such analysis for about 100 disc galaxies with cogent
success \cite{sm02}. Analysis of more galaxies with similar
success will further buttress the case for MOND. It will also make
it ever more unlikely that MOND is ``just some clever way of
summarizing tight relations between baryon and dark matter (DM)
distributions''. MOND predicts a unique and history independent
relation between the baryon distribution and the dynamics for each
individual galaxy. It is inconceivable that the DM doctrine will
reproduce this kind of predictions for individual galaxies: In the
DM picture the relation between the distribution of the
dissipative, interactive, and potentially explosive baryons and
that of the inert DM should depend drastically on the unknown
history of formation, intrinsic evolution, and interaction with
the exterior of each individual galaxy.
\par
Beside this quality that we can achieved by quantity, we should
also strive to extend the range of galaxy types that we subject to
rotation-curve analysis. For example, it is important to extend
the analysis to include more high accelerations
(high-surface-brightness) galaxies, for which MOND predicts
initially declining rotation curves, and of which there are only a
few examples currently in the literature. In another limit, that
of low mass (and hence low velocity) spirals, there exist MOND
analyses for quite a few galaxies, with particularly reliable
data, having maximum rotational velocities in the range of $50-100
\kms$ \cite{mb88,bbs91,br92,dbm98,sv98}.

For spirals of even lower masses it becomes increasingly difficult
to come by specimens with very reliable data. Among other problems
such spirals tend to be of irregular shape and kinematics,
kinematics that are more difficult to interpret as pure rotation
(e.g. with misaligned kinematic and photometric axes). They are also
characterized by increasing role of pressure support in comparison
with the rotational support: Dynamical study of such galaxies may
require then large asymmetric drift corrections, which are rather
uncertain. [see e.g., \cite{cote00,begum04a,begum04b}]. Obtaining
accurate distances is also an issue since the nature of these
galaxies constrain us to look at nearby ones for which Hubble
distances are not applicable and Cepheid distances are, most often,
unavailable.

Even so, these galaxies present a stringent test, not only because
of their extremeness, but also because they are by and large fully
in the deep MOND regime, showing very low accelerations at all
radii. Also, their mass is generically dominated by gas, whose
surface density distribution is determined directly and so the
comparison with MOND is, in many cases, rather insensitive to the
assumed or fitted M/L value. The MOND result in such cases is all
but a pure prediction as opposed to a one parameter fit.

Here we test MOND on three galaxies with maximum rotational
velocities of less than 50 kms$^{-1}$, which, according to MOND,
corresponds to baryonic masses below $4\cdot10^8 M_\odot$. These
have regular gas and velocity distributions. And, very importantly,
all three have high inclinations, which minimizes errors due to
inclination corrections. They also have relatively small asymmetric
drift corrections. We also describe the results for one galaxy that
is of even a lower mass, but isn't all that ideal from the above
points of view. We include it here more as a potential case study
for the possible effects of the uncertainties inherent in small
inclinations and large asymmetric-drift corrections.

Rotation curve predictions of MOND contain several sub predictions
pertaining to partial aspects of the dynamics. One of these is an
exact relation of the form $V^4_{\infty}=MG\a0$ between the
asymptotic rotational velocity, $V_{\infty},$ and the total
(baryonic) mass, $M,$ of a galaxy. This is not as probing a
prediction as a full rotation curve analysis but requires less
knowledge--e.g., of the exact mass distribution in a galaxy--and can
thus be applied to larger samples. The galaxies we consider here
have already featured in the mass-velocity plot shown by McGaugh
(2005,2006) and all indeed fall on the predicted MOND mass-velocity
relation, stretching its tested gamut to some five orders of
magnitude in mass.

We describe the data in section 2 and the results of MOND analysis
in section 3. These are discussed in section 4.

\section{The Data}

The data for the galaxies we consider come from recent work of
Begum, Chengalur, and coworkers. These include  NGC 3741
\cite{begum05}, KK98 250 (=UGC 11583) and KK98 251
\cite{begum04a}, and DDO 210 \cite{begum04b}. We do not analyze
two other low mass galaxies, GR8 \cite{begum03a} and CAM B
\cite{begum03b}, because we deem them unfit for RC analysis (see
reasons below).

NGC 3741 has an HI disk that is very extensive compared with the
light distribution, and that exhibits a well ordered velocity field.
It has an estimated distance of 3 \mpc based on the
tip-of-the-red-giant-branch method (TRGB), and its inclination
varies between $58\deg$ and $70\deg$ in a tilted-ring analysis.
Begum and Chengalur (2005) attribute the sharp rise of rotational
velocity at small radii to the bar (non-circular motions).

KK98 250 and KK98 251 also have well ordered velocity maps. The
first has an inclination of $80\deg\pm 4\deg$  based on the HI
ellipticity (consistent with optical one of $79\deg$). We use the HI
rotation curve from Begum and Chengalur (20004a) and the \Halpha
measurements in the inner parts from McGaugh, Rubin, \& de Blok
(2001), provided in tabular form by Stacy McGaugh. KK98 251 has an
inclination of $62\deg\pm 5\deg$ from the axes ratio of the HI
distribution (consistent with what is obtained  from the kinematics
in the tilted-ring fits: $65\deg$); we know of no \Halpha velocities
for this galaxy. There are no direct distance measures to these two
galaxies. The estimated distance to both, of 5.6 \mpc, is based on
group membership (the NGC 6946 group) and on a mean distance to the
group estimated from the brightest stars in eight members. The
estimated asymmetric-drift correction for these two galaxies, which
was included in the rotation curve we use, is small \cite{begum04a}.

The distance to DDO 210, of 0.95 \mpc, is also based on the TRGB.
The deduced rotation curve of DDO 210 is subject to two worrisome
uncertainties. Its estimated inclination is small: $27\deg\pm 7\deg$
as deduced from the ellipticity of the HI distribution which is
rather irregular (a kinematic inclination could not even be derived
from the velocity field). Also, while the estimated asymmetric drift
correction for DDO 210 is not as large as that for CAM B it does
require a substantial correction (see below).

The two galaxies from Begum and Chengalur we have not included are
GR8 and CAM B. The data for GR8 is practically impossible to
interpret as a rotating galaxy (e.g., its kinematic axis is almost
perpendicular to the major axis of the HI distribution) and so no
rotation curve is given.   CAM B presents more regular kinematics
than GR8, but the main problem we see in this case is that it is
subject to a very large (and very uncertain) asymmetric-drift
correction. So large, in fact, is the correction that it totally
dominates the resulting curve. Taking the correction applied by
Begum and Chengalur at face value it follows that CAM B is by and
large supported by random motions, with the rotational support
making only a small contribution. The deduced pressure force is 4
times the centripetal force at 0.1 kpc, with the ratio between the
two increasing to about 9 at the last measured point. The asymmetric
drift correction is not only uncertain, as always, but in this case
the standard formula used is not really valid.

\section{Results}

\begin{table}
\caption{Adopted and derived properties of sample galaxies}
\label{symbols}
\begin{tabular}{@{}lccccc}
\hline Object & $D(\mpc)$ & $L(10^8 L_\odot)$ & $M_{gas}(10^8
M_\odot)$ &
    $a_f$  & $(M/L)_M$ \\
\hline
KK98 250 & 5.6  & 1.15 (I)  & 1.6. & 0.2 & 1.5 \\
NGC 3741  & 3.0  &0.28 (B)  & 2.0 & 0.12 & 1.5 \\
KK98 251  & 5.6  &0.22 (I)  & 1.0 & 0.2 & 0.5 \\
DDO 210 & 0.95  & 0.027 (B)  & .036  & 0.19 & 0.2 \\
\hline
\end{tabular}

\medskip
The second column gives the distance as adopted by Begum,
Chengalur, and collaborators; the third column gives the
luminosity in either the I band or B band; the fourth column gives
the total gas mass including primordial helium, the fifth column
gives the centrepital acceleration at the last measured point on
the rotation curve in acceleration units of $10^{-8}\cmst$, and
the last column gives the deduced MOND M/L value in solar units,
for the band corresponding to the photometry.
\end{table}

\begin{figure}
 \plotone{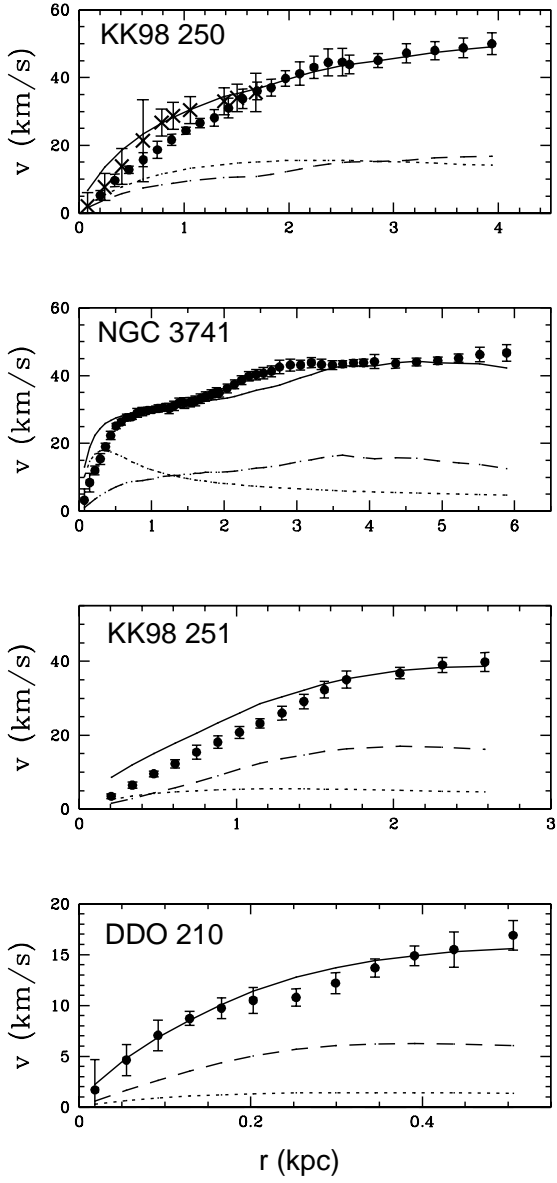}
 \caption{The MOND rotation curves (solid curves) for the four sample
galaxies assumed to be at the distances adopted by Begum et al. The
filled circles show the rotation curve obtained from the 21 cm line
observations. In the case of KK98 250, the crosses show the rotation
curve in the inner region derived from H$_\alpha$ observations. The
dotted and dashed curves show the Newtonian rotation curves for the
stars and the gas respectively.}
\end{figure}

To calculate the MOND curve $v(r)$ we use the algebraic MOND
relation
$$\mu(a/\a0)a=a_N$$
with $a=v^2/r$ the true (MOND) acceleration and $a_N$ the Newtonian
acceleration calculated from the mass distribution. We take
$\a0=10^{-8}\cmst$, and use $\mu(x)=x(1+x^2)^{-1/2}$. Note though
that the exact form of $\mu$ is rather immaterial here since all
four galaxies are deep in the MOND regime so only the linear, $x\leq
0.3$, part of $\mu$ is probed. Also, for such galaxies the exact
value of $\a0$ is not well constrained as its effect is degenerate
with that of the inclination and that of the distance, which is not
well known; the three appearing in the fits in the combination $\a0
D^2 sin^4i$ (Milgrom 1988).

As in previous analysis \cite{sv98} both the stars and gas are
assumed to be in a thin disk and the HI surface
density is increased by a factor of 1.3 to account for the presence
of primordial helium.  The distances (all as adopted by Begum,
Chengalur, and collaborators), luminosities, total gas mass,
acceleration at the last measured point of the rotation curve, along
with the fitted mass-to-light ratios, are given in Table 1.

Figure 1 shows the comparison of the measured rotation curve with
that deduced from MOND for the four galaxies under study. In the
case of KK98 250, the HI kinematic data has been supplemented with
\Halpha data in the central regions.

\section{Discussion}

Due to the paucity of data, as well as the intrinsic uncertainties
noted above, there have been very few studies of low rotation
velocity, gas dominated dwarfs in the literature. Lake (1989) had in
his sample two galaxies with nominally as low masses as ours, but
their stated velocities, and hence masses, were based on low and
uncertain inclinations (see discussion in Milgrom 1991).
C\^{o}t\`{e} Carignan \& Freeman (2000) discusses SDIG, a galaxy
with very low rotational velocities, but it is similar to GR8 and
not amenable to a decent rotation curve analysis.
\par
We show here, albeit with a very small sample, that, like
their higher mass kin, the lowest mass spirals with reasonable data
conform to the predictions of MOND. On the face of it, the agreement
is not as nearly perfect as in many higher quality specimens, but
considering the intrinsic uncertainties, the predicted rotation
curves appear to be quite reasonable.  The primary errors, not
present in higher quality samples, result from uncertain distance
determinations, non-symmetric or irregular gas distributions,
ambiguous inclinations, and highly uncertain asymmetric drift
corrections. Note also that here, as in all other data presented in
the literature, only errors on the measured rotation curves are
given, but there are no errors quoted on the mass distribution,
which must be reflected as errors in the predicted MOND rotation
curve.

We now discuss the results for individual galaxies.

KK98 250: The plotted rotation curve is from two sources: In the
inner region, the higher resolution H$_\alpha$ data going up to
about 2\kpc \cite{mcg01}, and also the lower resolution 21 cm
observations that extend twice as far \cite{begum04b}. In the inner
parts the agreement with the MOND prediction is clearly better for
the H$_\alpha$ data. This emphasizes the importance of higher
resolution data in the inner parts especially for such small, low
velocity galaxies. With the best fit MOND M/L value, the stars and
gas turn out to contribute similarly to the mass in this galaxy.

NGC 3741:  This is a strongly gas dominated galaxy, with
$M_{gas}/L\approx 7$; so that the adopted $M/L$ for the stellar disk
hardly matters-- the overall shape of the MOND rotation curve is
determined essentially by the gas distribution.  Indeed the
principal uncertainty in this predicted curve results from the
uncertainty of the azimuthally averaged gas density distribution. A
15\% larger distance for this galaxy (consistent with the distance
uncertainty) improves the agreement between the MOND and observed
curves.

 KK98 251:  The predicted
asymptotic rotation velocity agrees well with that observed. With
the adopted stellar M/L value the stars contribute only about 0.1 of
the total mass; so, the deduced asymptotic MOND velocity is
practically a prediction based only on the observed gas mass, and is
rather insensitive to the assumed M/L value (within reasonable
range). As for KK98 250, the MOND curve lies above the observed 21
cm points in the inner regions. For KK98 250 this is rectified by
the \Halpha data. The two galaxies are similar and we can expect, or
predict, that the \Halpha rotation curve,  which is still lacking
for KK98 251, will better agree with the MOND curve in the inner
parts.

DDO 210: This galaxy was already considered in light of MOND by
Begum and Chengalur (2004b) who found good agreement with MOND, but
leaving $\a0$ a free parameter they found a best fit value of
$\a0=1.7\cdot 10^{-8} \cmst$, which is higher than the standard
value. We notice that this is an artifact of a numerical error in
the fitting program used and their correct best-fit value is in fact
$\a0=0.85 \cdot 10^{-8} \cmst$, near to the value we use. We
included this galaxy in our analysis mainly for the opportunity to
demonstrate and discuss some of the sources of uncertainty besetting
the analysis. The deduction of the gravitational acceleration in the
plane of the galaxy (as expressed in terms of the published
``rotation curve'') still requires a large asymmetric drift
correction (although not as large as in CAM B): In the outer parts
the ``pressure'' force is 4 times larger than centripetal force. The
indicated errors stem only from the measured HI rotation curve and
do not include uncertainties in the asymmetric drift correction. In
addition, the adopted inclination is small and uncertain: The
velocity field is clearly distorted and the tilted ring inclination
fit failed to converge. For these reasons Begum and Chengalur
(2004b) assumed a constant inclination estimated from the
ellipticity of the HI distribution, which gave $i=27\pm7\deg$ (this
value totally disagrees with the optical inclination, which may be
dominated by patches). This galaxy has $M_{gas}/L\approx 1.3$. With
the stellar $M/L$ value we find of about 0.2, the stellar mass is
rather unimportant so the fit is not sensitive to $M/L$. However,
this value could be significantly higher than what we give,
especially considering the above mentioned uncertainties.

A large sample of dwarf galaxies has been collected by Swaters
(1999) and an analysis in terms of MOND is in preparation (Swaters
and Sanders 2006). The predicted MOND rotation curves agree, in
general, quite well with the observed curves particularly
considering that galaxies in this sample are subject to similar
uncertainties to those discussed here.

Note added: In a paper just posted on the net Gentile et al.
(astro-ph/0611355) present new data, analysis, and MOND fits for
NGC 3741. They find that to get a velocity curve from the velocity
data one must include radial motions of considerable magnitude.
Correcting for those is a rather uncertain, but necessary,
procedure. This is in line with the worries expressed above
regarding uncertainties in the analysis of such low mass galaxies.
With their corrected rotation curve, Gentile et al. find a rather
better agreement with MOND than we do, albeit for a somewhat
larger distance of about 3.5 Mpc, compared with the pre-fixed
value of 3 Mpc we use (our fit would also improve somewhat with a
larger distance).

 \acknowledgements
We are grateful to Ayesha Begum for providing data and for help in
tracking down the error in the standard fitting program.  We also
thank Stacy McGaugh for the \Halpha data on KK98 250 and for
comments on the manuscript.

\clearpage
\end{document}